\documentclass{PoS}
\def\lsim{\:\raisebox{-0.5ex}{$\stackrel{\textstyle<}{\sim}$}\:}
\def\etal{{\it et al.}}
\def\go{\rightarrow  }
\def\lra{\leftrightarrow}
\def\be{\begin{equation}}
\def\ee{\end{equation}}
\def\br{\begin{eqnarray}}
\def\er{\end{eqnarray}}
\def\brn{\begin{eqnarray*}}
\def\ern{\end{eqnarray*}}
\def\rf#1{{(\ref{#1})}}

\def\b {{\beta}}

\def\w {{\omega}}
\def\sss{\scriptscriptstyle}
\def\nn{\nonumber}
\def\ie{{\em i.e., }}

\def\M {{{\cal M}}}
\def\T {{{\cal T}}}
\def\R {{{\cal R}}}

\def\Ket#1{||#1 \rangle}
\def\Bra#1{\langle #1||}

\def\sss{\scriptscriptstyle}
\def\kb {{\bf k}}
\def\rb {{\bf r}}

\def\gV{g_{\mbox{\tiny V}}}
\def\gM{g_{\mbox{\tiny M}}}

\def\gP{g_{\mbox{\tiny P}}}
\newcommand{\Mass}{\mathrm{M}}
\def\gA{g_{\mbox{\tiny A}}}
\newcommand{\mass}{\mathrm{m}}

\def\ga{\overline{g}_{\mbox{\tiny A}}}

\def\gp{\overline{g}_{\mbox{\tiny P}}}

\def\gw{\overline{g}_{\mbox{\tiny W}}}

\def\mbs{\mbox{\boldmath$\sigma$}}
\def\mbn{\mbox{\boldmath$\nabla$}}

\def\absk {{|\kb|}}

\def\bit{\begin{itemize}}
\def\eit{\end{itemize}}
\def\bnu{\begin{enumerate}}
\def\enu{\end{enumerate}}
\def\lsim{\:\raisebox{-0.5ex}{$\stackrel{\textstyle<}{\sim}$}\:}

\title{Muon capture rates
within the projected QRPA}

\ShortTitle{Muon capture  within the PQRPA}

\author{\speaker{Danilo Sande Santos}%
\\
Departamento de Ci\^encias Exactas e Tecnol\'ogicas, Universidade
Estadual  de Santa Cruz,
CEP 45662-000 Ilhe\'us, Bahia-BA, Brazil
\\
E-mail: \email{danilosandesantos@hotmail.com}}
\author{{Arturo R. Samana}\\
Departamento de Ci\^encias Exactas e Tecnol\'ogicas, Universidade
Estadual  de Santa Cruz,
CEP 45662-000 Ilhe\'us, Bahia-BA, Brazil\\
 E-mail:  \email{arsamana@uesc.br}}
\author{{Francisco Krmpoti\'c}\\
Instituto de F\'isica La Plata, CONICET, Universidad Nacional de La Plata\\
E-mail: \email{krmpotic@fisica.unlp.edu.ar}}
\author{{Alejandro J. Dimarco}\\
Departamento de Ci\^encias Exactas e Tecnol\'ogicas, Universidade
Estadual  de Santa Cruz,
CEP 45662-000 Ilhe\'us, Bahia-BA, Brazil\\
 E-mail:  \email{atote@uesc.br}}

\abstract{ The conservation of the number of
 particles within the QRPA plays an important role in the evaluation muon  capture rates
  in all light nuclei with $A\lsim 30$.
The  violation of the CVC by the Coulomb field  in this mass region is of minor importance,
but this effect could be quite relevant for medium and heavy nuclei studied previously.
The extreme sensitivity of the muon capture rates on the $pp$ coupling strength
 in nuclei with large neutron excess when described within the QRPA is pointed out.
We reckon  that the comparison between theory and data for the
inclusive muon capture is not a fully satisfactory test on the nuclear model that is used.
The exclusive muon transitions are much more  robust for such a purpose.}

\FullConference{XXXIV edition of the Brazilian Workshop on Nuclear Physics\\
                 5-10 June 2011\\
                 Foz de Igua\c{c}u, Parana state, Brasil}

\begin{document}

\section{Introduction}

Among  different semileptonic processes, the muon capture
 is one
of the weak observables that, together with the $\beta$-decay, has
available a fruitful set of experimental data that were collected
in the last fifty years. Several works were focused to establish
the universal $V-A$ character of nuclear muon capture, the role of
induced currents, second-class currents, and nonexistence of $V+A$
interactions. It is known that the experimental value of the
induced pseudoscalar coupling $g_{\sss P}$ is the least known
 of the four constants ($g_{\sss V}, g_{\sss A}, g_{\sss M},g_{\sss P}$)
 defining the weak nucleon
current.

Its size is dictated by chiral symmetry arguments,
and its measurement represents an important test of quantum
chromodynamics at low energies~\cite{Fea97}. During the past two decades a large
body of new data relevant to the coupling $ g_{\sss P}$ has been accumulated
from measurements of radiative and non radiative
muon capture on targets ranging from $^{3}$He  to complex nuclei. Only
 transitions to unnatural parity states depend on $ g_{\sss P}$, as can be seen
from Eq. \rf{2.5}.
A summary of references on these issues are cited in review papers
Ref.~\cite{Mea01,Gor04,Gor06}.

Simultaneously, the muon capture processes have been used to scrutinize the
nuclear structure models, since they provide a testing ground for wave functions and, indirectly,
for the interactions that generate them. Being the momentum transfer of the order of the
muon mass $\mass_\mu= 105.6$  MeV, the phase space and the nuclear response favor
lower nuclear excitation energies, and thus  the transitions to  nuclear
states in the giant resonance region are the dominate ones.
We will cite only a few of them. Most of these works were done within the shell model (SM)
framework~\cite{Gor06,Hax90,War94,Vol00,Aue02}. Several studies were performed by
 employing the random phase approximation (RPA)~\cite{Vol00,Kol94,Kol94a,Zin06}.
In the last work, where  the total muon capture rates for  a large number
of nuclei with $6<Z<94$ have been evaluated, the authors claimed that an
important benchmark was obtained by introducing the pairing correlations.
They have done this ad-hoc by  multiplying the one-body transition
matrix elements by the BCS occupation probabilities.
However, we know that
the quasiparticle RPA (QRPA) formalism
is a full self-consistent procedure to describe consistently both i) short-range
particle-particle ($pp$) pairing correlations, and ii) long-range particle-hole ($ph$),
correlations handled with RPA. Quite recently, the relativistic
QRPA (RQRPA) was applied in the calculation of total muon capture rates
on a large set of nuclei from $^{12}$C to $^{244}$Pu, for which
experimental values are available~\cite{Mar09}.

In the present  work we do a systematic study of the muon
capture rates of nuclei with $12 \leq A \leq 56$ masses ($^{12}$C, $^{20}$Ne, $^{24}$Mg,
$^{28}$Si, $^{40}$Ar, $^{52}$Cr, $^{54}$Cr, $^{56}$Fe) within
the number projected  QRPA (PQRPA).
The motivation for this investigation comes from the successful description
of weak observables in the triad
$\{{{^{12}{\rm B}},{^{12}{\rm C}},{^{12}{\rm N}}}\}$ within this model~\cite{Krm02,Sam11}.
There, it was shown that the projection procedure  played an  essential role
in properly accounting for the configuration mixing in the  ground state wave
function of $^{12}$N.
The employment of PQRPA for the inclusive
$^{12}$C$(\nu_e,e^-)^{12}$N cross section, instead of the continuum
RPA (CRPA) used by the LSND collaboration in the analysis of
${\nu}_\mu \go{\nu}_e$ oscillations of the 1993-1995 data sample,
leads to an increased oscillation probability~\cite{Sam06}.
 The charge-exchange PQRPA,
derived from the time-dependent variational principle, was used to
study the two-neutrino $\beta\beta$-decay amplitude $\M_{2\nu}$ in $^{76}$Ge~\cite{Krm93}.
In that work, the projection procedure was less important and the
QRPA and PQRPA yield qualitatively similar results for $\M_{2\nu}$.
The PQRPA was recently used to calculate the
$^{56}$Fe$(\nu_e,e^-)^{56}$Co cross section~\cite{Sam08}.

We will also give a glance on the violation of the CVC by the Coulomb field, which
was worked out recently~\cite{Sam11}, and appears in the first operator \rf{2.4}
for natural parity states
\footnote{ When the consequences of the  CVC are not considered, as in Ref.~\cite{Gor04},
the factor $({\mass_\mu-\Delta E_{\rm
Coul}-E_B^\mu})/{E_\nu}$ in this relation goes to unity.}.
This effect is expected to be tiny for the nuclei studied here, since $\Delta E_{\rm Coul}$ is relatively small in comparison
with $\mass_\mu$; it goes from $3.8$ MeV in  $^{12}$C to $9.8$ MeV in $^{56}$Fe.
\section{$\mu$-capture rates formalism}
When negative muons pass through matter, they can
be captured into high-lying atomic orbitals. From there they then
quickly cascade down into the $1S$ orbit with binding energy  $E_B^\mu$,  where two competing
processes occur: one is ordinary decay $\mu^-\go
e^-+\nu_\mu+{\tilde \nu}_e$ with characteristic free lifetime $2.197\times 10^6$ sec, and the other is (weak) capture by the nucleus
$\mu^- + (Z, A)\go (Z-1, A)+ \nu_\mu$.  The latter, naively
 expected to scale with  $Z$, is drastically enhanced by an additional  factor of
 $Z^3$, originating from the square of the atomic wave function $\phi_{1S}$ evaluated at
 the origin [2]. Thus, its rate is roughly proportional to $Z^4$ and dominates decay at
 large $Z$. This dominance is however significantly diminished by the gradual decrease of the
  effective-charge correction   factor  $\R(Z)$  \cite{Mar09,Wal04}.

  Then
the muon capture rate from the ground state in the initial nucleus $(Z, A)$ to the
state ${\sf J}^\pi_n$ in the final nucleus  $(Z-1, A)$
 reads
 \br
 \Lambda({\sf J}^\pi_n)&=&\frac{E_\nu^2}{2\pi}|\phi_{1S}|^2\R(Z)
\T_{\Lambda}(E_\nu,{\sf J}^\pi_n), \label{2.1}\er
where
 \br
E_\nu\equiv\kappa=\mass_\mu-(\Mass_n-\Mass_p)-E_B^\mu-\w_{{\sf J}^\pi_n}
 \label{2.2}\er
 is the neutrino energy, and \br
\T_{\Lambda}(E_\nu,{\sf J}^\pi_n)&=&4\pi G^2
\left[|\Bra{{\sf J}^\pi_n}{\sf O}_{\emptyset{ {\sf J}}}(E_\nu)-
{\sf O}_{0{ {\sf J}}}(E_\nu)\Ket{0^+}|^2+2|\Bra{{\sf J}^\pi_n}{\sf O}_{-1{ {\sf J}}}(E_\nu)
 \Ket{0^+}|^2\right],
\label{2.3}\er
is the transition probability, being the Fermi
coupling constant $G=(3.04545\pm 0.00006){\times} 10^{-12}$ natural units.
The nuclear operators are:
\br
{\sf O}_{\emptyset{\sf J}}-{\sf O}_{0, \sf  J}&=& \gV\frac{\mass_\mu-\Delta E_{\rm
Coul}-E_B^\mu}{E_\nu}\M^{\sss V}_{\sf J},
\nn\\
{\sf O}_{-1{\sf J}} &=&-(\gA +\gw){\M}^{\sss A,I}_{-1{\sf J}}
+\gV \M^{\sss V,R}_{-1\sf J}, \label{2.4}\er
for  {\it natural parity states} ($\pi=(-)^J$,
i.e., $J^\pi=0^+,1^-,2^+,3^-,\cdots$),
and
\br
{\sf O}_{\emptyset{\sf J}}-{\sf O}_{0 \sf  J}&=&\gA\M^{\sss A}_{\sf J}
+ \left(\gA+\ga-\gp\right)\M^{\sss A}_{0{\sf J}},
\nn\\
{\sf O}_{-1{\sf J}} &=&-(\gA +\gw)\M^{\sss A,R}_{-1\sf J}
-\gV{\M}^{\sss  V,I}_{-1{\sf J}},
\label{2.5}\er
for {\it unnatural parity states} ($\pi=(-)^{J+1}$,  i.e., $J^\pi=0^-,1^+,2^-,3^+,\cdots$).
The elementary   operators are:
\br
\M^{\sss V}_{\sf J}=j_{\sf J}(\rho) Y_{{\sf J}}(\hat{\rb})&,&
\M^{\sss V}_{{m\sf J}}={\rm M}^{-1}\sum_{{\sf L}\ge 0}i^{ {\sf J-L}-1}
F_{{{\sf LJ}m}}j_{\sf L}(\rho)[ Y_{\sf L}(\hat{\rb})\otimes\mbn]_{{\sf J}},
\nn\\
\M^{\sss A}_{\sf J}=
{\rm M}^{-1}j_{\sf J}(\rho)Y_{\sf J}(\hat{\rb})(\mbs\cdot\mbn)&,&
\M^{\sss A}_{{m\sf J}}=\sum_{{\sf L}\ge 0}i^{ {\sf J-L}-1}
F_{{{\sf LJ}m}}j_{\sf L}(\rho)
\left[Y_{{\sf L}}(\hat{\rb})\otimes{\mbs}\right]_{{\sf J}}
\label{2.6}\er
where $F_{{\sf LJ}m}=(-) ^{1+ m}(1,-m{\sf J}m|{\sf L}0)$
is a Clebsch-Gordon coefficient, $\rho=\absk r$, and  the  superscripts
$ R$, and $I$ in \rf{2.4} and \rf{2.5} stand for real and imaginary
pieces of the operators \rf{2.6}. Moreover,
\br
\Delta E_{\rm Coul}\cong \frac{6e^2Z}{5R} \cong 1.45 ZA^{-1/3}~~\mbox{MeV},~~~~
E_B^\mu=(eZ)^2\frac{\mass_\mu}{2}\cong 2.66\times 10^{-5}Z^2{\mass_\mu},
\label{2.7}\er
and
\be
\ga=\gA\frac{E_\nu}{2\Mass},~~
\gw=(\gV+\gM)\frac{E_\nu}{2\Mass};~~\gp=\gP\frac{E_\nu}{2\Mass},
\label{2.8}\ee
with $g_{\sss V}$, $g_{\sss A}$, $g_{\sss M}$, and $g_{\sss P}$, being the effective
vector, axial-vector, weak magnetism, and pseudo-scalar
coupling constants, respectively.
 We adopt
 \br g_{\sss V}&=&1, ~~~g_{\sss A}=1.135 , ~g_{\sss M}=3.70,
~g_{\sss P}=g_{\sss A}\frac{2\Mass \mass_\mu
}{k^{2}+\mass_\pi^2}\cong 6.7, \label{2.9}\er where the value for
$g_{\sss P}$ comes from the  PCAC, pion-pole dominance and the
Goldberger-Trieman relation~\cite{Gol58}, and for $g_{\sss A}$ we
use the same value as in Ref.~\cite{Mar09}.

 From Eqs. (A6) and (A7) in Ref.~\cite{Sam11}
 one sees that  $g_{\sss P}$   is contained in
axial-vector pieces of both  operators  ${\sf O}_{\emptyset{\sf
J}}$ (temporal), and ${\sf O}_{0, \sf  J}$ (spacial). They
contribute destructively, being dominant the second one.
In Ref. \cite{Mar09} $g_{\sss P}$  appears only in the temporal operator.
However, after making use of the energy conservation condition
\rf{2.2}, \ie $\kappa\cong \mass_\mu+k_{\emptyset}$
($k_{\emptyset}=-\w_{{\sf J}^\pi_n}$) one ends up with the same result for
${\sf O}_{\emptyset{\sf J}}-{\sf O}_{0\sf  J}$.

The $0^+\lra0^-$ transitions are determined by two nuclear
matrix elements only: $\M^{\sss A}_{\sf 0}$ and $\M^{\sss A}_{0{\sf
0}}$, as can be seen from the first relation in \rf{2.5}. As such
they are
 the most appropriate to extract  the magnitude of  $g_{\sss P}$ from
 the muon capture experiments.
In fact,  studies
of the $^{16}{\rm O}(0^+_1) \go ^{16}{\rm N}(0^-_1)$ transition
 within large-basis SM calculations have yielded values of $g_{\sss P}= 6-9$
\cite{Hax90}, and $g_{\sss P}=7.5\pm0.5$
\cite{War94} that are consistent with the estimate \rf{2.9} as well as with
theoretical prediction $g_{\sss P}=8.2$ from
chiral symmetry arguments \cite{Fea97}.
More recently, Gorringe~\cite{Gor06} reported from the SM study of muon capture
$^{40}{\rm Ca}(0^+_1) \go ^{40}{\rm K}(0^-_1)$ have extracted from the experimental result
of $\Lambda$ the values $g_{\sss P}=14.3^{+ 1.8}_{-1.6}$, and $g_{\sss P}=10.3^{+ 2.1}_{-1.9}$.

\section{Numerical results}

For the set of nuclei discussed here we have  adopted the single-particle energies (s.p.e.)
from  the self-consistent calculation performed by Marketin
\etal~\cite{Mar09} within  the relativistic Hartree-Bogoliubov model
(RHB), using effective Lagrangians with density-dependent
meson-nucleon couplings and DD-ME2 parametrization.
The  residual interaction is approximated by the $\delta$-force
(in MeV fm$^3$)
\[
V=-4 \pi \left(v_sP_s+v_tP_t\right) \delta(r),
\]
with singlet ($v_s$), and triplet ($v_t$) coupling constants
  different $ph$, $pp$, and pairing
channels.
The  proton and neutron pairing parameters   $v_s^{pair}(p)$ and
$v_s^{pair}(n)$, used in  solving the BCS and PBCS equations,
 were determined  from the experimental data by  the adjusting
 procedure described in Ref.~\cite{Hir90}.
For the   parameters in the $ph$ channel we employe  the values
$v_s^{ph}=27$ and $v_t^{ph}=64$, which were obtained  in
 a systematic study of the GT resonances~\cite{Krm94}.
The ratio $t={v_t^{pp}}/{v_s^{pair}}$ was considered  as free
parameter within the $pp$ channel. It was found~\cite{Kor02}  that
the muon capture, just like the $\b\b$-decay, probes the final leg
of a $\b\b$-transition and as such  strongly depends on the
strength of the $pp$ interaction. Even worse, the QRPA model
collapses as whole in the physical region of
$t$~\cite{Krm93,Krm94,Krm93a}.
Yet, the distinction
between the initial and final legs in the $\b\b$-decay only makes
 sense in nuclei that possess an appreciable neutron excess, which doesn't
happen in nuclei under discussion where $N\cong Z$. Moreover, the results of
the PQRPA calculations in $^{12}$C, displayed in Fig. 5 of
Ref.~\cite{Krm05} suggest that the choice
$t=0$ could be appropriate for the description of  $N\cong Z$ nuclei.
Therefore, this value of the $pp$ coupling strength is adopted here.

\begin{figure}
\vspace{-3.cm}
\begin{center}
 \includegraphics[width=9.cm,height=12.cm]{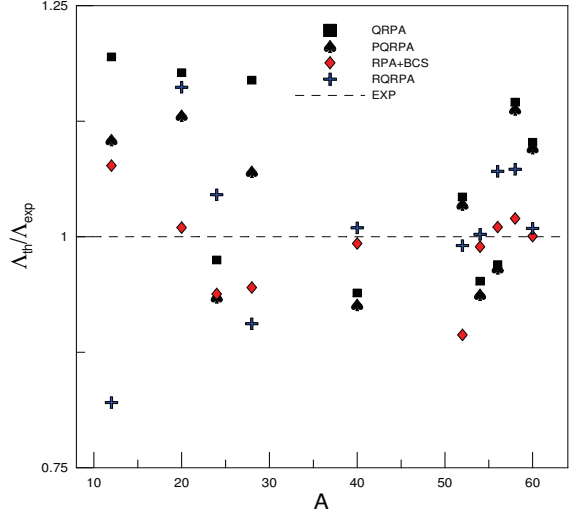}
\end{center}
\vspace{-3.cm} \caption{\label{figure1}(Color online)
Ratios of  theoretical to experimental inclusive muon capture rates for different
 nuclear models, as
function of the mass number $A$. The present QRPA and PQRPA results, as well as
the RQRPA calculation~\cite{Mar09}
were done with $\gA=1.135$, while in the
RPA+BCS model~\cite{Zin06} was  used the unquenched value $\gA=1.26$ for all multipole
operators, except for the GT ones  where it was reduced to $\gA\sim1$.}
\end{figure}

 Ratios of  theoretical to experimental inclusive muon capture rates for different
 nuclear models are exhibited in Fig. \ref{figure1}. It is self evident that the number
 projection plays an important role in light nuclei with $A \lsim 30$, making that the
 PQRPA agrees better with data than the plain QRPA. On the other hand it is difficult
 to judge whether our estimates are better or worse than the previous
 ones~\cite{Zin06,Mar09}.

 We have found that the consequences of the violation of the CVC by the Coulomb
 potential~\cite{Sam11} for the nuclei considered here is very tiny. In fact,
the major effect appears in $^{56}$Fe, where the total muon capture
is reduced  from $\Lambda=4260 \times 10^3$~s$^{-1}$ to $\Lambda=4056 \times 10^3$~s$^{-1}$.
\begin {table}[h]
\begin{center}
\caption{\label{table1} Energies (in units of MeV)
and exclusive muon capture rates (in units of $10^3$~s$^{-1}$) for
the bound excited states in $^{12}$B.
Besides the present PQRPA result, we also show a previous one~\cite{Krm02},
as well as those evaluated within  the SM~\cite{Aue02}, and the RPA~\cite{Kol94,Kol94a}.}
\label{table1}
\newcommand{\cc}[1]{\multicolumn{1}{c}{#1}}
\renewcommand{\tabcolsep}{0.3pc} 
\renewcommand{\arraystretch}{1.2} 
\bigskip
\begin{tabular}{| c c|c|c|c|c|c|}\hline
Model            &${\sf J}^\pi_n$&$1^+_1$&$2^+_1$&$2^-_1$&$1^-_1$&$\Lambda_{inc}$\\\hline\hline
PQRPA             &E         & $0.00$& $0.43$& $6.33$& $6.83$&\\ 
                  &$\Lambda$ &$8.80$& $0.20$& $0.60$& $0.85$&$37$\\
\hline
PQRPA~\cite{Krm02}&E         & $0.00$& $0.50$& $2.82$& $3.31$&\\
                  &$\Lambda$ & $6.50$& $0.16$& $0.18$& $0.51$&$40$\\
\hline
SM~\cite{Aue02}   &E         & $0.00$& $0.76$& $1.49$& $1.99$&\\
                  &$\Lambda$ & $6.0 $& $0.25$& $0.22$& $1.86$&\\\hline
RPA {\cite{Kol94,Kol94a}}  &$\Lambda$&$25.4~(22.8)$&$\leq
10^{-3}$&$0.04~(0.02)$&$0.22~(0.74)$&\\
\hline
Exp.~\cite{Mea01,Sto02}&E         & $0.00$& $0.95$& $1.67$& $2.62$&\\
                 &$\Lambda$ & $6.00\pm 0.40$& $0.21\pm 0.10$& $0.18\pm
0.10$& $0.62\pm 0.20$&$38\pm 1$\\
\hline\hline
\end{tabular}
\end{center}
\end {table}

In the case of $^{12}$C we have at our disposal also the experimental data
for exclusive muon capture rates to bound excited states
${\sf J}^\pi_n=1^+_1,2^+_1,2^-_1$, and $1^-_1$
in $^{12}$B~\cite{Mea01,Sto02}. They have been discussed
previously in the framework of the PQRPA~\cite{Krm02,Krm05}, but for
the sake of completeness we show them again in Table \ref{table1}.
The most relevant to highlight in this table is that, while both
PQRPA calculations of the inclusive muon capture rates agree fairly
well with the experiment, the corresponding exclusive reactions are
very different in the two calculations.
In other words, the agreement between theory and data for the
inclusive muon capture does not guarantee the goodness of the model that is used.

\section{Final remarks}

We have shown that, when the capture of muons
 is evaluated in the context of the QRPA, the conservation of the number of
 particles is very important not only for carbon but in all light nuclei with $A < 30$.
The consequence of this is the superiority of  the PQRPA on the  QRPA in this
nuclear mass region, as can be seen from Fig. \ref{figure1}.

The violation of the CVC by the Coulomb field  in this mass region is of minor importance, since in \rf{2.4} is
$\Delta E_{\rm Coul}+E_B^\mu$ is $\cong 11.7$ MeV, which is small in comparison with $\mass_\mu$.
However, this effect could be quite relevant for medium and heavy nuclei studied in
Refs.~\cite{Gor06,Zin06}. For instance, for $^{208}$Pb is $\Delta E_{\rm
Coul}+E_B^\mu\cong 39.0$ MeV, which implies a reduction
of the operator ${\sf O}_{\emptyset{\sf J}}-{\sf O}_{0, \sf  J}$ for
 natural parity states by a factor $0.37$, or equivalently that its contribution is only
 $\sim 13\%$ of that when the Coulomb field is not considered.

 We agree with the finding of  Kortelainen and  Suhonen~\cite{Kor02}
  on the extreme sensitivity of the muon capture rates on the $pp$ coupling strength
 when described within the QRPA,  as well as on a possible collapse of
 this approximation for the ${\sf J}^\pi_n=1^+_1$ state.
Yet, in our opinion  the  QRPA  behaves in this way   dominantly  in  nuclei with a
large neutron excess such as those analyzed in Refs.~\cite{Zin06,Mar09}.
It is clear that the RQRPA calculation~\cite{Mar09}  is sensitive to the $pp$ coupling, while the
 RPA+BCS model~\cite{Zin06} is not since it totally ignores the $pp$ interaction.

Finally, we conclude  that the comparison between theory and data for the
inclusive muon capture is not a fully satisfactory test on a nuclear model.
The exclusive muon transitions are much more  robust with respect to such a comparison.

\section*{Acknowledgements}
This work was partially supported by the Argentinean agency CONICET under
contract PIP 0377. A.R.S and D.S.S.  acknowledge the support by Brazilian agency FAPESB and
UESC, and   thank to Nils Paar for the values of s.p.e.
used in this work. D.S.S thanks to CPqCTR, where the numerical calculations were
performed.

\end{document}